\documentclass{article}

    \PassOptionsToPackage{numbers, compress}{natbib}
\usepackage[final]{nips_2017}

\usepackage[utf8]{inputenc} % allow utf-8 input
\usepackage[T1]{fontenc}    % use 8-bit T1 fonts
\usepackage{hyperref}       % hyperlinks
\usepackage{url}            % simple URL typesetting
\usepackage{booktabs}       % professional-quality tables
\usepackage{amsfonts}       % blackboard math symbols
\usepackage{nicefrac}       % compact symbols for 1/2, etc.
\usepackage{microtype}      % microtypography
\usepackage{adjustbox}
\usepackage[skip=-1pt]{caption}
\usepackage{graphicx}
\usepackage{verbatim}
\usepackage{amsmath}
\usepackage{enumitem}

\title{Speech Enhancement for Virtual Meetings on Cellular Networks}

\author{
\adjustbox{max width=\textwidth}{
\begin{tabular}{cccc}
\centering
Hojeong Lee & Minseon Gwak & Kawon Lee & Minjeong Kim \\& Joseph Konan & Ojas Bhargave\\
\end{tabular}}\\
Carnegie Mellon University\\ 
Pittsburgh, PA 15213\\
\{hojeongl, mgwak, kawonl, minjeong, jkonan,  obharga\}@andrew.cmu.edu
}

\begin{document}

\maketitle

\begin{abstract}
We study speech enhancement using deep learning (DL) for virtual meetings on cellular devices, where transmitted speech has background noise and transmission loss that affects speech quality. Since the Deep Noise Suppression (DNS) Challenge dataset of \textit{Interspeech 2020} does not contain practical disturbance, we collect a transmitted DNS (t-DNS) dataset using Zoom Meetings over T-Mobile network. We select two baseline models: Demucs and FullSubNet.
The Demucs is an end-to-end model that takes time-domain inputs and outputs time-domain denoised speech, and the FullSubNet takes time-frequency-domain inputs and outputs the energy ratio of the target speech in the inputs.

The goal of this project is to enhance the speech transmitted over the cellular networks using deep learning models.
\end{abstract}

\section{Introduction}

Speech enhancement (SE) has been widely studied for various edge devices and as preprocessing steps for various automatic systems \citep{gannot2017consolidated}.
In particular, as remote work using virtual meetings with cellular devices becomes more common, SE for the mobile meeting applications is essential.

The classical SE was driven by signal processing methods, such as Wiener filtering and spectral subtraction \citep{boll1979suppression, scalart1996speech}. 
However, recent studies have revealed the efficiency of data-driven methods, including deep learning \citep{lu2013speech, xu2014regression, weninger2014single, weninger2015speech, zhao2018convolutional}.

\begin{figure}[b!]%
    \centering
    \includegraphics[width=1\textwidth]{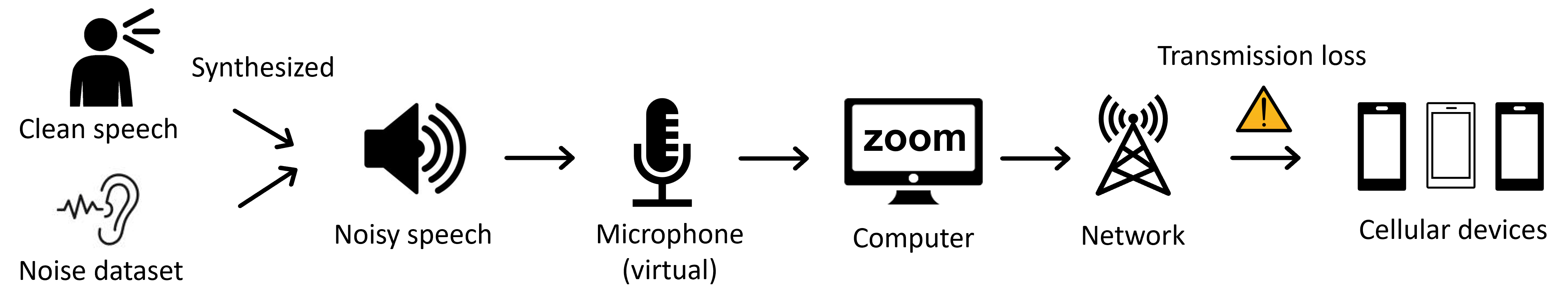}
    \caption{Data acquisition process of t-DNS}
    \label{fig:intro}
\end{figure}

The Deep Noise Suppression (DNS) Challenge dataset of \textit{Interspeech} 2020 has been released for data-driven SE research \citep{reddy2020interspeech}.
Recent DL-based SE studies have been conducted with the DNS dataset \citep{weninger2015speech, chen2017long, li2020online, hao2021fullsubnet}. However, the DNS dataset does not reflect the effect of transmission loss in the real-world network communication process.

In this project, we newly collect a transmitted DNS (t-DNS) dataset through the process shown in Figure \ref{fig:intro}.
The t-DNS data set contains data traversed by T-mobile network.
We aims to propose deep learning models that enhance the speech in the t-DNS dataset scoring better than \textit{`auto'} mode of Zoom's built-in background noise suppression model in terms of perceptual metrics and acoustic metrics.
With two baseline models, Demucs \citep{defossez2020real} and FullSubNet \citep{hao2021fullsubnet}, we introduce an auxiliary loss in terms of acoustic metrics known as the extended Geneva Minimalistic Acoustic Parameter Set (eGeMAPS) \citep{eyben2015geneva} to make the eGeMAPS features well-preserved in the denoised speech. 
To the best of our knowledge, we are the first to propose the SE dataset and model for virtual meetings over cellular networks in the real world. 
When all data processing work is completed, it is expected that the t-DNS dataset and model will be published online and used for future SE studies.

\section{Literature Review}

Recent studies on SE have contributed to enhancing the perceptual quality of denoised speech~\citep{zhao2018perceptually, hsieh21_interspeech, yang2022improving}.

The methods to obtain great perceptual quality are divided into metric-based learning and feature-based learning.
The metric-based learning aims to train a model that outputs denoised speech that results in good evaluation when a certain perceptual metric is calculated with target speech.
\citep{zhao2018perceptually} used an auxiliary loss of Short Time Objective Intelligibility (STOI).

The feature-based learning aims to train a model that outputs denoised speech, which has similar features to the clean or target speech.
This can be achieved by designing a loss function that captures the divergence between the target and denoised speech with respect to features of interest.
\citep{hsieh21_interspeech} proposed a phone-fortified
perceptual loss to use the phonetic information in speech in training models.
\citep{yang2022improving} proposed an auxiliary eGeMAPS loss to prevent the output speech from being distorted compared to the target speech in regard to eGeMAPS features.

\section{Model Description}
We select two baseline models: Demucs \citep{defossez2020real} and FullSubNet \citep{hao2021fullsubnet}.
The main difference is that the Demucs is an end-to-end model while the FullSubNet is a separate learning model.
The Demucs takes a raw time-domain waveform input and outputs denoised speech, which is also in the time domain.
By contrast, the FullSubNet takes a time-frequency-domain input and outputs values to compose the final denoised speech, which requires pre- and post-processing models for inputs and outputs.

\subsection{Demucs}
In the Demucs \citep{defossez2020real}, noisy speech $\mathbf{x}\in\mathbb{R}^T$ is considered as the sum of the clean speech $\mathbf{s}\in\mathbb{R}^T$ and noise $\mathbf{n}\in\mathbb{R}^T$ as follows:
\begin{align}
    \mathbf{x}=\mathbf{s}+\mathbf{n}.
\end{align}
The Demucs model $f$ is trained so that $f(\mathbf{x})=\hat{\mathbf{s}}\approx \mathbf{s}$.
The architecture of the SE model consists of a multi-layer convolutional encoder-decoder network with a sequence modeling LSTM network, which transforms the latent output of the encoder into a nonlinear transformation.
The model $f$ is trained with two types of loss functions: time-domain and time-frequency-domain losses.
The time-domain loss $L_\text{time}$ is the L1 loss between the clean and denoised output speech of $f$, i.e.,
\begin{align}
    L_\text{time}=\frac{1}{T}\vert\vert\mathbf{s}-\hat{\mathbf{s}}\vert\vert_1.
\end{align}
The time-frequency-domain loss $L_\text{T-F}$ consists of the spectral convergence loss $L_\text{sc}$ and magnitude loss $L_\text{mag}$, i.e., $L_\text{T-F}=L_\text{sc}+L_\text{mag}$, where
\begin{align}
    L_\text{sc}&=\frac{\vert\vert\vert S\vert-\vert\hat{S}\vert\vert\vert_F}{\vert\vert\vert S\vert\vert\vert_F},
    \\
    L_\text{mag}&=\frac{1}{T}\vert\vert\log\vert S\vert-\log\vert \hat{S}\vert\vert\vert_1,
\end{align}
with $S$ and $\hat{S}$ are the short-time Fourier transform (STFT) of $\mathbf{s}$ and $\hat{\mathbf{s}}$, respectively.
Moreover, multiple time-frequency-domain losses can be used with respect to different STFT resolution for the number of fast Fourier transform bins, hop sizes, and lastly window lengths.

The end-to-end property of the Demucs is beneficial in transfer learning and in that less domain knowledge is required to use the Demucs.
The performance of the causal/noncausal Demucs with proper data augmentation skills, such as reverbing with two sources and partial dereverberation, reached state-of-the-art models in both objective and subjective measures. 
Also, the Demucs enhanced automatic speech recognition systems without retraining on noisy conditions.

\subsection{FullSubNet}
The FullSubNet \citep{hao2021fullsubnet} is a fusion model of SE models that independently utilize fullband and subband information on short-time Fourier transform (STFT) of speech data.
Fullband models take the full band, up to the Nyquist frequency, of the STFT data and capture the global cross-band spectral characteristics of input speech.
By contrast, subband models take data in a partial frequency band and model local spectral patterns with fewer model parameters than fullband models.

As in the time domain, the STFT of noisy speech can also be represented with the STFT of the clean speech and noise, as follows: 
\begin{align}
    X = S + N,
\end{align} 
where $X$, $S$, and $N$ are the STFT of $\mathbf{x}$, $\mathbf{s}$, and $\mathbf{n}$, respectively.
Let $A(t,f)$ denote the $(t,f)$th component of an STFT matrix $A\in\mathbb{C}^{T\times F}$, where $T$ is the number of frames and $F$ is the number of frequency bins of STFT.
In the FullSubNet architecture, the fullband LSTM model $g_{\text{fullband}}$ takes $\mathbf{X}_t=\left[\vert X(t,0)\vert,\vert X(t,1)\vert,\cdots,\vert X(t,F-1)\vert\right]^T\in\mathbb{R}^F$ as an input and extracts the fullband feature.
The subband LSTM model $g_{\text{subband}}$ takes an augmented input, which is the concatenation of the fullband output and subband spectra, i.e., $\left[\vert X(t,f-N)\vert,\cdots,\vert X(t,f)\vert,\cdots,\vert X(t,f+N)\vert, g_{\text{fullband}}(X)\right]^T\in\mathbb{R}^{2N+2}$, and predicts the complex ideal ratio mask (cIRM), $M(t,f)\in\mathbb{C}$, which measures the energy ratio of the target speech to the entire noisy input speech for each time-frequency bin, i.e., $S(t,f) = M(t,f) * X(t,f)$.
The real and imaginary parts of the cIRM, $(M_r, M_i)$, for $X(t,f)=X_r+iX_i$ and $S(t,f)=S_r+iS_i$ is defined as follows \citep{williamson2015complex}:
\begin{align}
    M_r=K\tanh(\frac{C}{2}\cdot\frac{X_rS_r+X_iS_i}{X_r^2+X_i^2}), \\
    M_i=K\tanh(\frac{C}{2}\cdot\frac{X_rS_i-X_iS_r}{X_r^2+X_i^2}),
\end{align}
where $K$ and $C$ are hyperparameters.
The ground truth cIRM, $M$, can be calculated from the clean and noisy speech pair, and the final denoised speech can be constructed from the output cIRM values and input noisy speech.
Thus, the FullSubNet model $g$ is trained so that $g(X)=\hat{M}\approx M$.
The $g$ is trained with $L_{\text{cIRM}}$, which measures the mean squared error between the true and estimated cIRMs.
It is shown in \citep{hao2021fullsubnet} that the FullSubNet outperforms state-of-the-art models on the DNS dataset, and the information obtained in the fullband and subband models is complementary.

\section{Dataset}
The t-DNS dataset will be created based on the DNS Challenge dataset \citep{reddy2020interspeech}. The DNS Challenge dataset aims to provide an extensive and representative dataset to train the speech enhancement models. It contains 500 hours of clean speech from 2,150 speakers and a noise data set with at least 500 clips for 150 audio classes. Also, it contains test data with and without reverberation, and we will focus on the test data without the reverberation. Noisy speech is generated by synthesizing clean and noise speech data. The synthesized noisy speech is then sent across a virtual microphone, Zoom Meetings, T-mobile network and finally to cellular devices, as shown in Figure \ref{fig:intro}. In the Zoom Meetings, a low-level built-in noise suppression model will be used to minimize the impact of the speech enhancement with using it. The data sent to each cellular device is collected by the computer through the audio interface.

\section{Evaluation Metric}
This section introduces the metrics for estimating the performance of our SE model. We explain target metrics utilizable in our project. All three metrics are classified as relative metrics, which require a reference signal to compare a given signal.

\begin{itemize}[leftmargin=*]
\item \textbf{Frequency weighted Segmental Signal to Noise Ratio (fwSegSNR)}
\vspace{.5em}\\
Time-domain and frequency-weighted measurements, Signal to Noise Ratio (SNR) and fwSegSNR are both based on a clean signal $X$ enhanced signal $\hat{X}$. This is given a different weight for each frequency. $W(j,m)$ is the weight on the frequency band of $j$th, and $K$ is the number of bands. $M$ is the total number of frames in the signal. $X(j,m)$ is critical critical band magnitude of clean signal at $m$th frame, $j$th frequency frequency band.

\begin{align} %\label{eqn:my_equation}
\mathrm{fwSegSNR}=\frac{10}{M}\sum^{M-1}_{m=0}\frac{\sum^K_{j=1}W(j,m)\log_{10}\frac{X(j,m)^2}{(X(j,m)-\hat{X}(j,m))^2}}{\sum^K_{j=1}W(j,m)}
\label{stoi eq1}
\end{align}

\end{itemize}

\begin{itemize}[leftmargin=*]
    \item \textbf{Perceptual Evaluation of Speech Quality (PESQ)}
    \vspace{.5em}\\
    PESQ performs well in a wide range of codecs and network conditions. The core part consists of aggregating the disturbance to measure the audible error in three steps each by using $p$ norm as Equation (\ref{pesq eq}); frame-by-frame disturbance, split second disturbance, and speech length averaged disturbance. The notation $N$ in Equation (\ref{pesq eq}) indicates the total number of data in each norm-calculating part. PESQ returns a mean opinion score (MOS) from 0 to 5, with higher scores indicating better quality. Usually, PESQ indicates WB-PESQ, a wide band PESQ, and NB-PESQ indicates a narrow band PESQ. WB-PESQ, which has the benefit of transferring higher data rates, reads the input signal with input filter of 2 while NB-PESQ, which has the benefit of better sensitivity and range, does it as 1.

    \begin{align}
    L_p = \left(\frac{1}{N}\sum_{m=1}^{N} \text{disturbance}[m]^p\right)^\frac{1}{p}
    \label{pesq eq}
    \end{align}
\end{itemize}
\begin{itemize}[leftmargin=*]
\item \textbf{Short-Time Objective Intelligibility (STOI)}
\vspace{.5em}\\
STOI is a function to calculate the linear correlation coefficient of clean speech and denoised speech data. In Equation (\ref{stoi eq1}), $X$ indicates a decomposed clean speech, and $Y$ is a decomposed noisy speech after DFT-based 1/3 octave band decomposition. In Equation (\ref{stoi eq1}), $d$ means the correlation coefficient of $X$ and $Y$ corresponds to each frame $m$ and one-third octave band $j$. In Equation (\ref{stoi eq2}), This $d$ is averaged as a single scalar value indicating the voice intelligibility, where $M$ represents the total number of frames and $J$ represents the number of one-third octave bands.

\begin{align} %\label{eqn:my_equation}
        d_j(m)=\frac{\sum_n\bigg(X_j(n)-\frac{1}{N}\sum_lX_j(l)\bigg)\bigg(Y_{j'}(n)-\frac{1}{N}\sum_lY_{j'}(l)\bigg)}{\sqrt{\sum_n\bigg(X_j(n)-\frac{1}{N}\sum_lX_j(l)\bigg)^2\sum_n\bigg(Y_{j'}(n)-\frac{1}{N}Y_{j'}(l)\bigg)^2}}
\label{stoi eq1}
\end{align}
\begin{align}
    d = \frac{1}{JM} \sum_{j,m}d_j(m)
\label{stoi eq2}    
\end{align}
\end{itemize}

\section{Loss Function}
\subsection{Temporal Acoustic Parameter Estimator}
As a training boost, we fine-tune the two baseline models with temporal acoustic parameter (TAP) loss. This aims to minimize the temporal divergence between clean and enhanced acoustic parameters. Its availability in both time domain and time-frequency domain enables us to use it to both Demucs and FullSubNet models.

For a given signal $\mathbf{y}$, let $\mathbf{A}_{\mathbf{y}} \in \mathbf{R}^{T \times 25}$ indicate the 25 temporal acoustic parameters in T discrete time frames, and $A_{\mathbf{y}}(t, p)$ indicate it by each parameter $p$ and discrete time frame $t$. Then, TAP parameter gives an estimate of $\mathbf{A}_{\mathbf{y}}$ as $\hat{\mathbf{A}}_{\mathbf{y}}$ as in Equation (\ref{TAP-eq1}).

\begin{equation}
\hat{\mathbf{A}}_{\mathbf{y}}=\mathcal{T} \mathcal{A} \mathcal{P}(\mathbf{y})
\label{TAP-eq1}
\end{equation}

TAP estimator is obtained from a pretrained recurrent neural network which minimizes the mean absolute error defined as Equation (\ref{TAP-eq2}). 

\begin{equation}
\operatorname{MAE}\left(\mathbf{A}_{\mathbf{y}}, \hat{\mathbf{A}}_{\mathbf{y}}\right)=\frac{1}{T P} \sum_{t=0}^{T-1} \sum_{p=0}^{P-1}\left|A_{\mathbf{y}}(t, p)-A_{\hat{\mathbf{y}}}(t, p)\right| \in \mathbb{R}
\label{TAP-eq2}
\end{equation}

Using TAP estimators makes end-to-end learning possible by overcoming the non-differentiable properties of acoustic parameters. 

\subsection{Temporal Acoustic Parameter Loss}

Temporal acoustic parameter loss, $\mathcal{L}_{\mathrm{TAP}}$, minimizes divergence between each TAP estimators of the clean and enhanced speech. The mathematical term is expressed in in Equation (\ref{TAP-eq3}). $\sigma(\boldsymbol{\omega})$ indicates the smoothed energy weights that emulates human hearing with bounded scales. Our loss function is a combination of the L1 loss and the acoustic loss. We control the weight of the acoustic loss by a parameter $\alpha$.

\begin{equation}
\mathcal{L}_{\mathrm{TAP}}(\mathbf{s}, \hat{\mathbf{s}})=\operatorname{MAE}(\mathcal{T} \mathcal{A}(\mathbf{s}) \odot \sigma(\boldsymbol{\omega}), \mathcal{T} \mathcal{A} \mathcal{P}(\hat{\mathbf{s}}) \odot \sigma(\boldsymbol{\omega}))
\label{TAP-eq3}
\end{equation}

\section{Experiments}

\subsection{Metric evaluation}
Table \ref{table:evaluation_trans} summarizes the results of 150 noise data in the DNS 2020 dataset after speech enhancement. We inserted the raw waveform form into the processes of Demucs and FullSubNet without additional training. Both methods show high speech enhancement performance. However, in the case of FullSubNet, the performance is better than that of Demucs in PESQ metrics, and in the rest of the metrics, the performance of Demucs is better.

\begin{table}[h]
\caption{Evaluation on enhancement in denoised speech compared to noisy speech}
\centering
%\resizebox{\textwidth}{!}
{% 
\begin{tabular}{l|ccccccc}
\hline
           & WB-PESQ  & STOI(\%)     & fwSNRseg(dB) \\ \hline
Noisy      & 1.582  & 91.51  & 12.62   \\ \hline
Demucs     & 2.647  & 96.52  & 17.17   \\
FullSubNet & 2.888  & 96.41  & 16.96   \\ \hline
\end{tabular}}
\label{table:evaluation_trans}
\end{table}

\subsection{Acoustic improvement}
In addition to speech-level metric evaluation of denoised speech, the acoustic parameters-improving abilities of the SE models were analyzed.
We used 25 acoustic parameters defined in the eGeMAPS.
The acoustic parameters include frequency-related parameters, energy or amplitude-related parameters, spectral balance parameters, and temporal parameters.
We denote the $i$th acoustic parameter vector of speech $\mathbf{u}$ as $A_\mathbf{u}^{(i)}\in\mathbb{R}^{T_\mathbf{u}}$ for $i=1,2,\cdots,25$, where $T_\mathbf{u}$ is the total number of time frames of $\mathbf{u}$.
To consider all denoised speech of an SE model $m$, let $\mathbf{A}_m^{(i)}$ be the augmented acoustic parameter vector such that $\mathbf{A}_m^{(i)}=\left[(A_\mathbf{u}^{(i)})^T\right]^T_{\mathbf{u}\in \mathcal{S}_m}\in\mathbb{R}^T$, where $\mathcal{S}_m$ is the set of all denoised speech of $m$ and $T=\sum_{\mathbf{u}\in \mathcal{S}_m}T_\mathbf{u}$.
For better interpretation, augmented acoustic parameter vectors were standardized with some specific mean $\mu_i$ and standard deviation $\sigma_i$ values obtained in a large speech dataset for each acoustic parameter $i$, as follows:
\begin{align}
    \mathbf{A}_m^{(i)}=\frac{\mathbf{A}_m^{(i)}-\mu_i\mathbf{1}_T}{\sigma_i},
\end{align}
where $\mathbf{1}_T$ is $T$-dimensional all-ones vector.

\begin{figure}[t!]%
    \centering
    \includegraphics[width=1\textwidth]{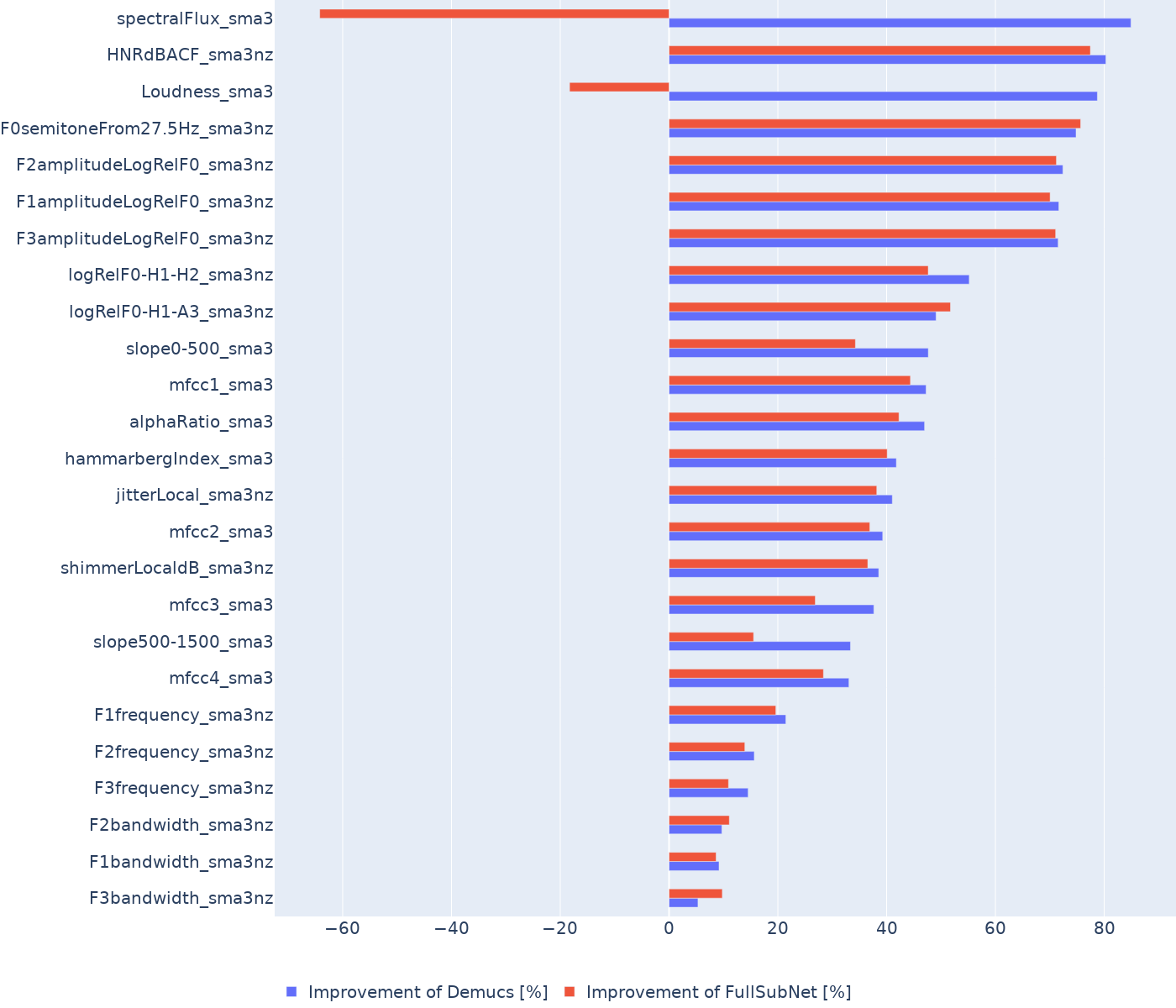}
    \caption{Improvement in the acoustic parameters of the denoised speech of two SE models}
    \label{fig:lld}
\end{figure}

\begin{figure}[t!]%
    \centering
    \includegraphics[width=1\textwidth]{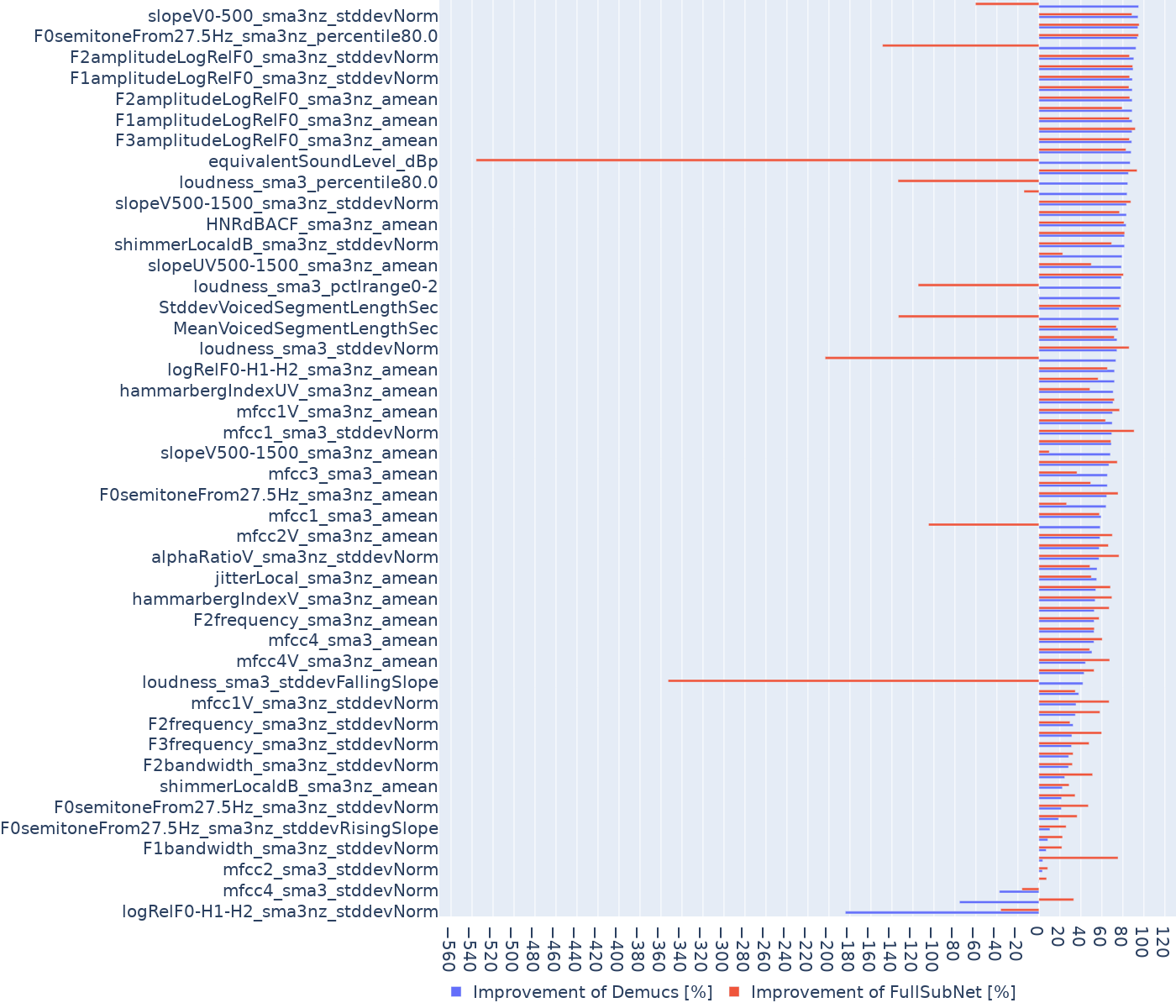}
    \caption{Improvement in the statistics of acoustic parameters of the denoised speech of two SE models}
    \label{fig:functionals}
\end{figure}

To see the acoustic improvement of $m$, We first evaluate the mean absolute error (MAE) of the acoustic parameter of denoised speech to the acoustic parameters of the corresponding clean speech for every $i$th parameter, as follows:
\begin{align}
    \text{MAE}_m^{(i)} = \frac{1}{T} \sum_{t=1}^{T} \left\vert \mathbf{A}_\text{clean}^{(i)}(t)-\mathbf{A}_m^{(i)}(t)\right\vert,
\end{align}
where $\mathbf{A}_m^{(i)}(t)$ is the $t$th component of $\mathbf{A}_m^{(i)}$ and $\mathbf{A}_\text{clean}^{(i)}$ denotes the augmented acoustic parameter vector of clean speech.
We then evaluated the acoustic improvement $I_m^{(i)}$ of an SE model $m$ for the $i$th acoustic parameter, as follows:
\begin{align}
    I_m^{(i)}=\frac{\text{MAE}^{(i)}_\text{noisy} - \text{MAE}^{(i)}_{m}}{\text{MAE}^{(i)}_\text{noisy}}\times 100,
\end{align}
where $\text{MAE}^{(i)}_\text{noisy}$ denotes the MAE for noisy speech.
The acoustic improvement in Demucs and FullSubNet, i.e., $I_\text{Demucs}$ and $I_\text{FullSubNet}$, is shown in Fig. \ref{fig:lld}, where the $y$-axis represents the 25 acoustic parameters in the eGeMAPS.
Moreover, the improvement with respect to the statistics for each acoustic parameter is also evaluated in Fig. \ref{fig:functionals}.
The SE models improved almost all acoustic parameters, as shown in Figs. \ref{fig:lld} and \ref{fig:functionals}.
Some acoustic parameters, such as `spectralFlux\_sma3' and `Loudness\_sma3', however, are degraded by the FullSubNet, which requires further analysis of the denoised speech of the FullSubNet.

\section{Results}
\subsection{Perceptual Evaluation}
\begin{table}
\caption{Metrics of speech enhancement quality}
\centering
% \resizebox{\textwidth}{!}{% 
\begin{tabular}{l|ccccccc}
\hline
           & fwSNRseg(dB) & PESQ & STOI(\%) \\ \hline
Clean       & - & - & - \\
Noisy      & 12.629  & 1.582  & 91.52   \\ 
Noisy Relay (Low)      & 4.804  & 1.549  & 79.76   \\ \hline
Industrial Denoising (Auto)      & 5.636  & 1.701  & 81.06   \\ \hline
Demucs (Baseline)     & 5.611  &  1.375 & 76.51   \\
Demucs (Fine-tuned)     & 6.772  &  1.397 & 80.18   \\ 
Demucs (Ours)     & 8.959  &  1.557 & 84.52   \\ \hline
FullSubNet (Baseline) & 5.712  & 1.511  & 78.2   \\ 
FullSubNet (Fine-tuned) & 6.546  & 1.496  & 80.27   \\
FullSubNet (Ours) & 8.897  & 1.631  & 84.01   \\ \hline
\end{tabular}
    \vspace{.5em}\\

\label{table2}
\end{table}

Table \ref{table2} shows the evaluation of each model in three metrics: fwSNRseg, PESQ and STOI. \textit{Noisy} is the raw noisy data before entering zoom. \textit{Noisy Relay} indicates the speech transmitted through zoom with 'low' mode of built-in background noise suppression. \textit{Industrial Denoising} indicates the speech transmitted through zoom with 'auto' mode of built-in background noise suppression. For each Demucs and FullSubNet, the three different models are used. \textit{Baseline} model is same as the provided model from the paper. \textit{Fine-tuned} model is further trained model with the training data. As the higher metrics means the better speech, there exists a degradation due to the transmission loss. The metrics of fine-tuned Demucs are better than those of the built-in low noise suppression model. However, the best Demucs model is worse than the auto mode in terms of PESQ for now. This is because the hyperparameter tuning is currently in progress. When the hyperparameter working is done, the metrics will be get much better.
The results from FullSubNet show similar trends to those from Demucs.

\subsection{Acoustic Evaluation}

\begin{figure}[h]%
    \centering
    \includegraphics[width=1\textwidth]{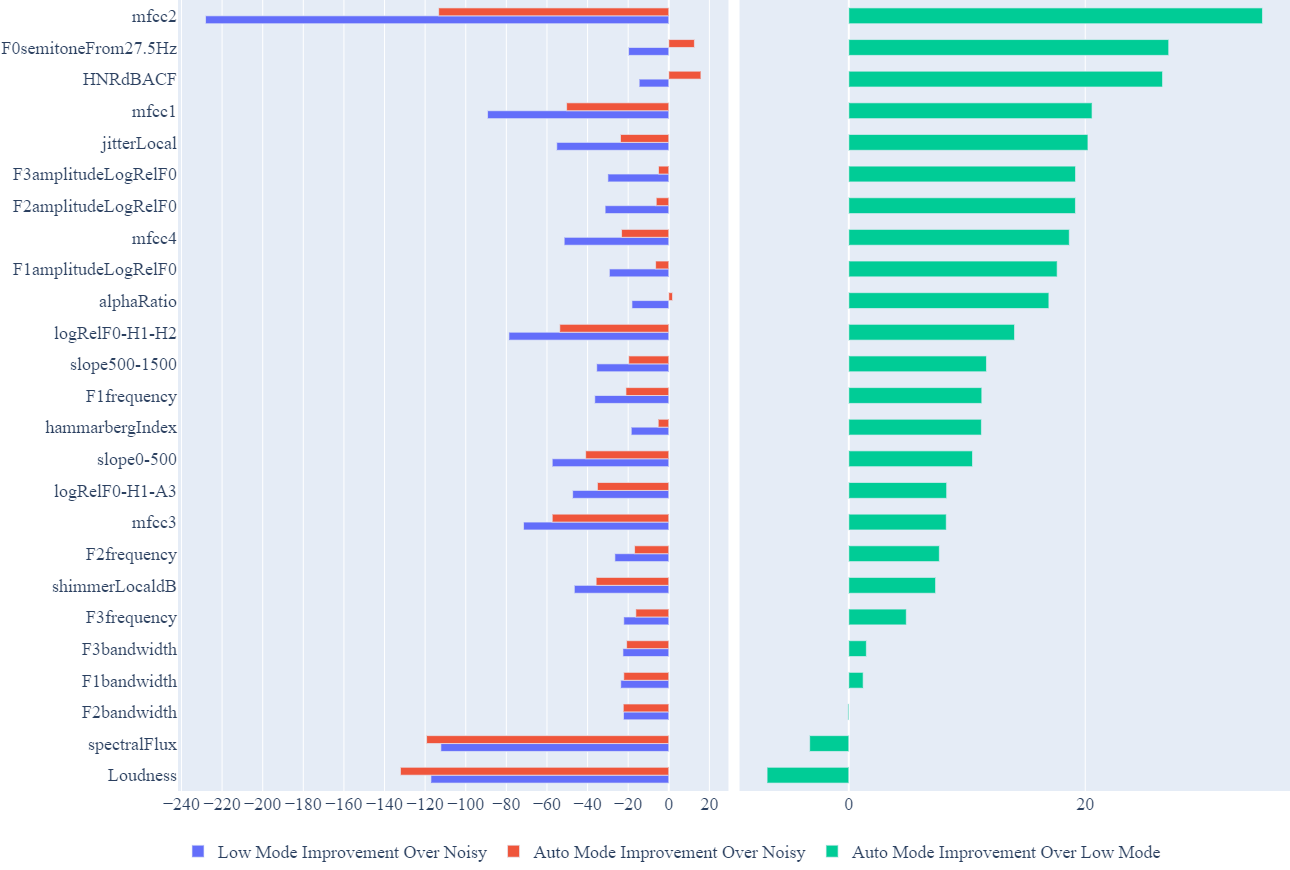}
    \caption{Acoustic parameters of Zoom built-in denoising modes}
    \vspace{.5em}
    \label{fig:eGeMAPS_Zoom}
\end{figure}

\begin{figure}[h]%
    \centering
    \includegraphics[width=1\textwidth]{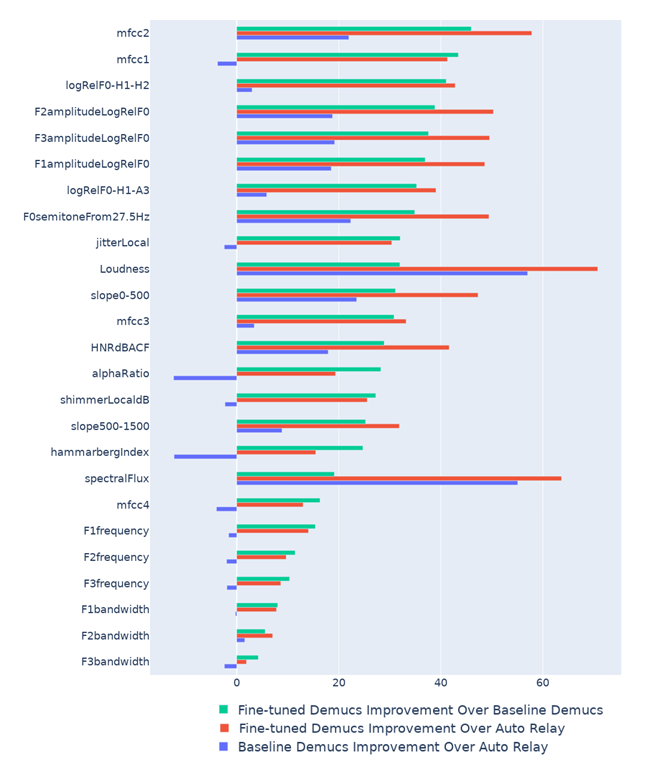}
    \caption{Acoustic parameters for Demucs}
    \vspace{.5em}
    \label{fig:egemaps_demucs}
\end{figure}

The improvement of acoustic metrics is measured as how well the input noisy speech is processed into enhanced speech. 
The left portion of Figure \ref{fig:eGeMAPS_Zoom} is the improvement of each Zoom's Low and Auto modes over untransmitted noisy speech. The right portion is about the improvement of auto mode over the low mode which shows that the auto mode is more powerful than the low mode. Even when using the Zoom's built-in noise suppression, noise added to the speech due to transmission on cellular networks degrades its speech in almost all aspects of eGeMAPS. In Figure \ref{fig:egemaps_demucs}, y-axis is for 25 acoustic parameters. The green and red bars represent the improvement of baseline and our models, repectively, compared to the Zoom's auto denoising mode. The blue bar represents how much our model is better than the baseline.
Our model showed better improvements in most of the acoustic parameters.

\section{Future Works}
As the rest of the dataset is being processed, we only can investigate the dataset transmitted through T-moblie network.
Once processing is done on the other 3 networks, we will compare the data from each of the 4 network provider and use SE to make the worst one the best.
Also, we are considering to analyze the acoustic characteristics of t-DNS. Then, we can optimize the acoustic parameters using characteristics that will make the noisy speech even better than the enhanced speech in this project.

\section{Conclusion}
The main contribution of our work is that we provide the t-DNS dataset which reflects the effect of transmission loss in the real-world cellular network communication. Also, we applied temporal acoustic loss function to fine tune the two baseline models, Demucs and FullSubNet. Our model beats the baseline models and the industrial denoised model, showing the effect of training on t-DNS dataset and temporal acoustic loss function.

\label{gen_inst}
\section{Division of work}
The project work was evenly distributed, and all team members participated in report writing and regular meetings throughout the semester.
\begin{itemize}
    \item \textbf{Hojeong Lee}: Results analysis, metric evaluation
    \item \textbf{Minseon Gwak}: FullSubNet implementation and experiments
    \item \textbf{Kawon Lee}: Demucs implementation and experiments, presentation
    \item \textbf{Minjeong Kim}: Results analysis, metric evaluation
\end{itemize}

\section{Github repository}
https://github.com/Minseon-Gwak/Speech-enhancement-zoom-phone

\bibliographystyle{plain}

\newpage

\section{Appendix}
Here are some example of our work: \href{https://cmu.app.box.com/s/p2s03iaf70xynom3y7kvv5u4i7v6fgy4}{FullSubNet Speech Enhancement Demo}

\end{document}